\documentclass{appolb}
  \usepackage{graphicx}
  \usepackage{epsfig}
  \usepackage{amssymb}
  \usepackage{subfigure}
  \usepackage{axodraw}

\def\({\left(}
\def\){\right)}

\newcommand{\be}{\begin{equation}}
\newcommand{\ee}{\end{equation}}
\newcommand{\nn}{\nonumber}
\newcommand{\bea}{\begin{eqnarray}}
\newcommand{\eea}{\end{eqnarray}}
\newcommand{\bfig}{\begin{figure}}
\newcommand{\efig}{\end{figure}}
\newcommand{\bc}{\begin{center}}
\newcommand{\ec}{\end{center}}
\newcommand{\bd}{\begin{displaymath}}
\newcommand{\ed}{\end{displaymath}}

\newcommand{\sintheff}{\sin^2\theta_{\mbox{eff}}^{\mbox{lept}}}
\newcommand{\itcaption}[2][]{\caption[#1]{\it #2}}

\begin{document}
\pagestyle{plain}

%%%%%%%%%%%%%%%%%%%%%%%%%%%%%%%%%%%%%%%%%%%%%%%%%%%%%%%%%%%%%%%%%%%%%
%%%%%%%%%%%%%%%%%%%%%%%%%%%%%%%%%%%%%%%%%%%%%%%%%%%%%%%%%%%%%%%%%%%%%

\begin{titlepage}
\nopagebreak
{\flushright{
        \begin{minipage}{38mm}
        WUE-ITP-2005-015
        Freiburg-THEP 05/13
        {\tt hep-ph/0512026}\\
        \end{minipage}        }

}
\vspace*{-1.5cm}                        
\vskip 3.5cm
\begin{center}
\boldmath
{\large \bf{Three-loop electroweak corrections \\[1mm]
to the $\rho$-parameter, $\sintheff$ and       \\[1mm]
the $W$-boson mass in the large Higgs mass limit}}\footnote{
Presented at the XXIX conference
`Matter to the Deepest', Ustron (Poland), September 2005.
Published in Acta Phys.\ Polon.\ B {\bf 36} (2005) 3275,
\copyright Acta Physica Polonica}
\unboldmath
\vskip 1.2cm
{\textsc{Radja Boughezal}
%\footnote{Email: {\tt Radja.Boughezal@physik.uni-wuerzburg.de}}
}\\[1mm]
{\it Institut f\"ur Theoretische Physik und Astrophysik, 
Universt\"at W\"urzburg,
\mbox{Am Hubland},
D-97074 W\"urzburg, Germany}\\[1.5mm]
and\\[1.5mm]
{\textsc{Bas Tausk}
% \footnote{Email: {\tt Tausk@physik.uni-freiburg.de}}
}\\[1mm]
{\it  Fakult\"at f\"ur Mathematik und Physik, 
Albert-Ludwigs-Universit\"at Freiburg, \\
D-79104 Freiburg, Germany}
\\[2mm] 
\end{center}
\vskip 1.5cm
\begin{abstract}
A review of the calculation of the leading three-loop electroweak
corrections to the $\rho$-parameter, $\sintheff$ and the $W$-boson
mass is presented. The heavy Higgs mass expansion and the renormalization
are discussed.
\vskip .7cm 
\flushright{
        \begin{minipage}{12.3cm}
{\it PACS}: 12.15.Lk, 14.80.Bn
        \end{minipage}        }
\end{abstract}
\vfill
\end{titlepage}

%%%%%%%%%%%%%%%%%%%%%%%%%%%%%%%%%%%%%%%%%%%%%%%%%%%%%%%%%%%%%%%%%%%%%
%%%%%%%%%%%%%%%%%%%%%%%%%%%%%%%%%%%%%%%%%%%%%%%%%%%%%%%%%%%%%%%%%%%%%
\section{Introduction}
\label{sec:intro}

The Electroweak Standard Model (EWSM) is in good agreement
with all experimentally known phenomena of electroweak origin,
with the exception of the evidence of neutrino mixing.
The only ingredient predicted by this model that has not been 
seen yet is the Higgs particle. The direct search at LEP provided
us with a lower limit on the mass of the Standard Model Higgs-boson
$m_H$ excluding the region below $114.4~$GeV~\cite{Barate:2003sz}.
The global fit of the experimental data to the Standard Model,
based on confronting theoretical predictions for electroweak
observables with their experimental values, favours a light 
Higgs-boson ( $m_H \leq 186~$GeV one-sided $95\%$ Confidence Level 
(CL)~\cite{LEP}).\\
However, there is a discrepancy of about $3.2\,\sigma$ between the two
most precise measurements of the electroweak mixing angle $\sintheff$,
which is used to set stringent bounds on the Higgs-boson mass. 
The measurement based on the leptonic asymmetry parameter $A_l$ at SLD,
together with the $W$-boson mass measurement from Tevatron and LEP, 
point to a light Higgs-boson with a mass slightly below the lower 
bound from the direct searches. On the other hand,
the measurement based on the $b$-quark forward-backward asymmetry 
$A_{FB}^{0,b}$ at LEP favours a relatively heavy Higgs-boson with 
a mass around $500~$GeV~\cite{LEP}. 
Since the center of mass energies of present colliders do not
allow to probe the region of a heavy Higgs-boson, the sensitivity 
of radiative corrections to low energy electroweak observables
to a heavy-Higgs boson mass becomes an important tool in setting
limits on $m_H$~\cite{LEP}.\\
\noindent As the Yukawa couplings are very small, the Higgs dependent
effects are limited to corrections to the vector-boson propagators.
They give rise to shifts, e.g., in the $\rho$-parameter,
the $W$-boson mass, and in the effective leptonic
weak mixing angle $\sintheff$.
These shifts are often parametrized by S, T and U or 
$\epsilon_1,\,\epsilon_2, \, \epsilon_3$~\cite{Eidelman:2004wy}.\\
\noindent For a light Higgs-boson, the Higgs mass dependence 
of theoretical predictions is mainly due to one-loop radiative
corrections to the gauge-boson propagators which grow logarithmically
with $m_H$~\cite{Veltman:1976rt}. However, because the Higgs 
self-interaction is proportional to $m_H^2$, higher order radiative 
corrections which grow like powers of $m_H$ could become important
if the Higgs-boson mass is much larger than the Z-boson mass.\\
\noindent At the two-loop level, the leading corrections are proportional
to $m_H^2$, but the numerical coefficient of these terms turns out 
to be very small~\cite{vanderBij2, Barbieri:1993ra}, 
and therefore they are not important for $m_H$ less than a few TeV.
However it has been suggested that the smallness of the two-loop
corrections may be somewhat accidental~\cite{akhoury}, therefore 
important effects might first appear only at the three-loop level.\\
The situation was only clarified recently by an explicit leading
three-loop calculation for three precision variables, 
the $\rho$-parameter, the electroweak mixing angle $\sintheff$, 
and the $W$-boson mass $M_W$~\cite{Boughezal:2004ef-2005}.
\\      

\noindent The electroweak $\rho$-parameter is a measure of the relative 
strengths of neutral and charged-current interactions 
in four-fermion processes at zero momentum transfer. In the Standard
Model, at tree level, it is related to the W and Z boson masses by:
\begin{equation}
\label{eq:rhotree}
\rho = \frac{M_W^2}{c_W^2\,M_Z^2} = 1 \, ,
\end{equation}
where $c_W = \cos \theta_W$. Including higher order corrections 
modifies this relation into
\begin{equation}
\rho = \frac{1}{1 - \Delta\rho}.
\end{equation}
Here $\Delta\rho$ parametrises all higher loop corrections which
are sensitive to the existence of a heavy Higgs particle.
The leading one- and two-loop corrections, which grow logarithmically
and quadratically with $m_H$ respectively, have been 
calculated~\cite{Veltman:1976rt, vanderBij:1983bw}.\\
\noindent The sine of the effective leptonic weak mixing angle
$\sintheff$ is defined in terms of the couplings of the $Z$-boson
to leptons. The complete electroweak fermionic corrections 
to $\sintheff$ at the two-loop level 
are known~\cite{Awramik:2004ge-Hollik:2005va}.
Recently, the Higgs-dependent electroweak two-loop bosonic
contributions to this observable have been 
completed~\cite{Hollik:2005ns}.   
For the $W$-boson mass, both the fermionic and the bosonic corrections
have been calculated at the two-loop level~\cite{Awramik:2003rn}.\\
\noindent In this contribution we review the leading three-loop 
bosonic corrections to the $\rho$-parameter, $\sintheff$ and 
$M_W$~\cite{Boughezal:2004ef-2005}, which grow like $m_H^4$ in
the large Higgs mass limit.\\
The calculation is organized in such a way that the leading
contributions come only from self energy corrections to
the gauge boson propagators, often referred to
as {\em oblique} corrections in the literature, and not from
vertex or box diagrams. This is achieved by our choice of
renormalization scheme. We should
mention at this point that the renormalization is performed up 
to the two-loop level only, removing sub-divergences
from the gauge boson self-energies, but not yet the overall
divergences. This is due to the fact that the three-loop
counter terms cancel in $\Delta^{(3)}\rho$, 
$\Delta^{(3)}\sin^2\theta_{\mbox{eff}}^{\mbox{lept}}$
and $\Delta^{(3)} M_W$.
The renormalized self-energies are then related to physical 
observables through the formalism of S, T and U parameters,
which was developed by Peskin and Takeuchi to describe the effect
of heavy particles on electroweak precision 
observables~\cite{Peskin}. They are defined in terms of 
the transverse gauge boson self-energies at zero momentum transfer
and their first derivative w.r.t their momentum. These self-energies
are not observable individually and may still contain 
ultraviolet divergences. However the three combinations:
\begin{eqnarray}
\label{eq:Sdef}
S &\equiv& \frac{4 s_W^2 c_W^2}{\alpha}
  \left( \Sigma^{\prime ZZ}_T
       - \frac{c_W^2-s_W^2}{c_W s_W} \Sigma^{\prime AZ}_T
       - \Sigma^{\prime AA}_T \right)
\\
\label{eq:Tdef}
T &\equiv& \frac{1}{\alpha M_W^2}
  \left( c_W^2 \Sigma^{ZZ}_T
             - \Sigma^{WW}_T
  \right)
\\
\label{eq:Udef}
U &\equiv& \frac{4 s_W^2}{\alpha}
  \left( \Sigma^{\prime WW}_T
       - c_W^2 \Sigma^{\prime ZZ}_T
       - 2 c_W s_W  \Sigma^{\prime AZ}_T
       - s_W^2 \Sigma^{\prime AA}_T \right)
\end{eqnarray}
where $s_W=\sin\theta_W$ and $\alpha$ is the fine structure constant,
are finite and observable.
Using this formalism, the shifts to $\rho,\,\sintheff$ and $M_W$
are parametrised as:
\begin{eqnarray}
\label{eq:deltarho}
&&\Delta\rho = \alpha \, T \,, \nn\\
\label{eq:deltamw}
&&\Delta M_W =
\frac{\alpha M_W}{2(c_W^2-s_W^2)}
\left( - \frac{1}{2} S + c_W^2 T + \frac{c_W^2-s_W^2}{4s_W^2} U \right) \,,
\nn\\
\label{eq:deltasin}
&&\Delta\sintheff \,=\,
\frac{\alpha}{c_W^2-s_W^2}
\left( \frac{1}{4} S - s_W^2 c_W^2 T \right) \,.
\end{eqnarray}
%
%%%%%%%%%%%%%%%%%%%%%%%%%%%%%%%%%%%%%%%%%%%%%%%%%%%%%%%%%%%%%%%%%%

\section{Calculation and renormalization}
The bare gauge boson self-energies are decomposed into 
transversal and longitudinal components according to
\begin{equation}
\Sigma^{X}_{\mu\nu}(p)
 = \left(g_{\mu\nu}-\frac{p_{\mu}p_{\nu}}{p^2}\right)
   \Sigma^{X}_T(p^2)
 + \frac{p_{\mu}p_{\nu}}{p^2} \,
   \Sigma^{X}_L(p^2) \, ,
\end{equation}
where $X = AA, AZ, ZZ, WW$.
The scalar functions $\Sigma^{X}_{T}(p^2)$ are then expanded in
a Taylor series in their momentum $p$ up to order $p^2$.
Higher order terms in $p^2$ are suppressed by powers of 
the heavy Higgs boson mass. For the $T$ parameter, only the constant
term of this expansion (i.e $p^2 = 0$) is required. After this
step, we are left with vacuum integrals which, in general,
depend on three different scales: $m_H,\,M_W$ and $M_Z$.
In order to separate the dependence of these integrals on
the large scale $m_H$ from the small scales $ M_W$ and $M_Z$
and extract the leading $m_H$ terms, we perform an asymptotic large
mass expansion following the method of the expansion 
by regions~\cite{smirnovbook}.   
The expansion is constructed by considering different regions in loop
momentum space, distinguished by the set of propagator momenta which are
large or small in those regions. In each region, a Taylor expansion of
all propagators in the small masses and in the small momenta of that
region is performed. Typically, the expansion generates
extra scalar products of loop momenta in the numerator
and higher powers of denominators, as compared to the original
diagrams. The resulting expression is then integrated over the whole 
loop momentum space. For the three-loop vacuum topology shown
in Fig~\ref{mercedes}, there are $15$ regions in loop momentum space
to consider (see~\cite{Boughezal:2004ef-2005} for more details). 
The distribution of large and small masses in each diagram decides
the number of regions that contribute to the corresponding integral
up to the leading terms we are interested in. For example, only
four regions give a non vanishing contribution to the diagram shown
in Fig~\ref{WWEXPREGexample}, up to $m_H^4$ order.
They correspond to:
\begin{itemize}
\item the all internal momenta large region. In this case one expands
the propagators in $M_{\phi}$ only, the result is a one-scale
three-loop integral.
\item the region where $k_1$ is small. Here we expand in $M_\phi$
and $k_1$, which leads to the product of a one-loop integral depending 
on $M_{\phi}$ times a two-loop integral that depends on $m_H$. 
\item the region where $k_1+k_2$ is small. After expanding in 
$M_{\phi}$ and $k_1+k_2$ we get, similarly to the previous case,
a product of one- and two-loop integrals.   
\item the region where $k_1+k_2+k_3$ is small. Here again we expand
in $M_\phi$ and $k_1+k_2+k_3$ and get a product of one- and 
two loop integrals. 
\end{itemize}
All the other regions produce either scaleless integrals, which are
zero in dimensional regularization, or terms which do not have
enough powers of $m_H$ to contribute to the leading order.
\begin{figure}[tb]
   \begin{center}
    \begin{picture}(80,80)(0,0)
    \SetColor{Black}
    \CArc(50,50)(40,0,180)
    \SetColor{Black}
    \CArc(50,50)(40,180,360)
    \SetColor{Black}
    \Vertex(50,90){1.5} \Vertex(50,50){1.5}
    \Line(50,90)(50,50)
    \Vertex(16,30){1.5}
    \Line(16,30)(50,50)
    \Vertex(84,30){1.5}
    \Line(84,30)(50,50)
%    \Text(55,70)[]{3}
%    \Text(5,50)[]{4}
%    \Text(95,50)[]{6}
%    \Text(50,17)[]{1}
%    \Text(39,35)[]{2}
%    \Text(61,35)[]{5}
%    \Text(50,0)[]{$I_{6}$}
    \end{picture}
    \end{center}
\vspace{-6mm}
    \itcaption{The vacuum three-loop six-propagators 
     topology.} 
     \label{mercedes}
\end{figure}
\noindent From the expansion in the $15$ regions we get two kinds of integrals:
\begin{itemize}
\item factorizable diagrams, which are products of one-(two) loop
vacuum integrals depending on $m_H$ and two-(one-) loop vacuum integrals
depending on $M_W$ and $M_Z$.
\item non-factorizable three-loop vacuum integrals, which are either
single-scaled depending on $m_H$, or double-scaled depending 
on $M_W$ and $M_Z$. 
\end{itemize}
\begin{figure}[tb]
   \setlength{\unitlength}{1.9pt}
%   \begin{picture}(130,130)(0,25)
    \begin{picture}(110,110)(0,25)
    \SetScale{1.9}
    \SetColor{Black}
    \GCirc(100,100){30}{1}
    \SetColor{Black}
    \Line(100,100)(100,130)
    \Line(75,84)(100,100)
    \Line(100,100)(125,84)
    \Photon(60,37)(90,72.5){2}{6}  \Vertex(90,72.5){1}
    \Photon(140,37)(110,72.5){2}{6}  \Vertex(110,72.5){1}
    \Text(59,30)[]{\small{$W^+$}}
    \Text(141,30)[]{\small{$W^+$}}
    \Text(100,64)[]{\small{H}}
    \Text(74,73.5)[]{$\phi^{-}$}
    \Text(128,73.5)[]{$\phi^{-}$}
    \Text(78,133)[]{$\phi^{-}$}
    \Text(128,133)[]{$\phi^{-}$}
    \Text(94,115)[]{\small{H}}
    \Text(80,95)[]{\small{H}}
    \Text(120,95)[]{\small{H}}
    \Text(100,74)[]{\small{$k_1$}}
    \Text(90,79)[]{\small{$k_1-p_1$}}
    \Text(110,79)[]{\small{$k_1-p_1$}}
    \Text(91,90)[]{\small{$k_2$}}
    \Text(107,90)[]{\small{$k_2+k_3$}}
    \Text(106,115)[]{\small{$k_3$}}
    \Text(150,116)[]{\small{$k_1+k_2+k_3-p_1$}}
    \Text(58,116)[]{\small{$k_1+k_2-p_1$}}
    \Text(82,45)[]{\small{$p_1$}}
    \SetColor{Blue}
    \LongArrow(68,38)(80,52)
    \SetColor{Blue}
    \LongArrow(120,52)(132,38)
    \end{picture}
\itcaption{An example of a three-loop $W$ self-energy diagram on which
the expansion by regions was applied. $H$ and $\phi^{-}$ refer to the Higgs
and Goldstone fields respectively.}
\label{WWEXPREGexample}
\end{figure}
The Integration-By-Parts (IBP) method~\cite{ibp}
is then used to reduce all the vacuum integrals to the master ones. 
We classify the single-scaled three-loop non-factorizable 
integrals into ten different kinds, depending on the distribution 
of masses in the propagators.
Their reduction to a small set of master 
integrals~\cite{Broadhurst:1998rz, Fleischer:1999mp} was done in two
ways. On the one hand, reduction formulae based on 
the Integration By Part identities have been constructed. 
On the other hand the Automatic Integral Reduction package 
$AIR$~\cite{Anastasiou:2004vj} was used as a cross check.
The latter was also used to reduce the double-scaled three-loop
non-factorizable integrals. Explicit formulae for their 
master integrals are not needed, as they all canceled once
we have summed over all diagrams.\\
\noindent The longitudinal parts of the gauge boson self-energies
are related to the self-energies of the Goldstones and mixings between
gauge bosons and Goldstones by a set of Ward identities. We have
verified that these Ward identities are satisfied by the full 
(i.e including tadpoles) unrenormalised self-energies up to order $p^2$.

As we have mentioned earlier, leaving out vertex and box contributions
requires a proper way of renormalizing. Our renormalization conditions
are fixed in such a way that the renormalization removes all
the terms of order $m_H^2$ and $m_H^4$ from the one- and two-loop
gauge boson self-energies, the charged and neutral Goldstone
self-energies and the mixings between gauge bosons and Goldstones.
This ensures that no two- or three-loop vertex or box graphs containing
such self-energies as subgraphs can give corrections that grow
like $m_H^2$ or $m_H^4$ in the large Higgs mass limit 
(see Fig.~\ref{vertexZZSE}).    
\begin{figure}
\begin{center}
\begin{picture}(100,70)(0,-5)
                 \SetColor{MidnightBlue}
                 \Vertex(30,0){1.5}
                 \Vertex(30,60){1.5}
                 \Vertex(60,30){1.5}
                 \Vertex(85,30){1.5}
                 \ArrowLine(100,0)(85,30)
                 \ArrowLine(85,30)(100,60)
                 \ArrowLine(0,0)(30,0)
                 \ArrowLine(30,0)(60,30)
                 \ArrowLine(60,30)(30,60)
                 \ArrowLine(30,60)(0,60)
                 \Photon(60,30)(85,30){3}{5}
                 \Vertex(30,20){1.5}
                 \Vertex(30,40){1.5}
                 \Photon(30,60)(30,40){-3}{3.5}
                 \Photon(30,0)(30,20){3}{3.5}
                 \CCirc(30,30){10}{MidnightBlue}{Gray}
%                {OliveGreen}
                 \Text(98,-4)[]{\Black{$\nu_{e}$}}
                 \Text(98,63)[]{\Black{${e^{-}}$}}
                 \Text(2,-6.5)[]{\Black{$\overline\nu_{\mu}$}}
                 \Text(2,66)[]{\Black{$\mu^{-}$}}
                 \Text(22,50)[]{\Black{\tiny $Z$}}
                 \Text(22,12)[]{\Black{\tiny $Z$}}
                 \Text(29,29)[]{\Black{\tiny{$2L$}}}
                 \Text(72,22)[]{\Black{\tiny{$W^{-}$}}}
              \end{picture}
\itcaption{An example of a three-loop vertex diagram which would contribute
to the leading terms if $\;\Sigma_{ZZ}^{(2L)}$ $\sim m_H^4$.}
\label{vertexZZSE}
\end{center}
\end{figure}
As a check on the renormalization, we have verified that 
the renormalized longitudinal photon self-energy and photon-Z mixing are zero.
%%%%%%%%%%%%%%%%%%%%%%%%%%%%%%%%%%%%%%%%%%%%%%%%%%%%%%%%%%%%%

\section{Results and conclusion}

The shifts to the electroweak precision observables $\rho$, $\sintheff$
and $M_W$ relative to their tree level values, expressed in terms
of $\alpha,\, G_F $ and $M_Z$, are given by
\begin{eqnarray}
&& \rho = \frac{1}{1 - \Delta\rho}\,,\\
&&  \sintheff =
\Delta\sintheff +
\frac{1}{2}
 -\sqrt{\frac{1}{4} - \frac{\pi\alpha}{\sqrt{2}\,G_F M_Z^2}}\, ,\\
&&
M_W = \Delta M_W +
 M_Z \sqrt{\frac{1}{2}
 +\sqrt{\frac{1}{4} - \frac{\pi\alpha}{\sqrt{2}\,G_F M_Z^2}}} \, ,
\end{eqnarray}
with $\Delta\rho$, $\Delta\sintheff$ and $\Delta M_W$ defined 
in~(\ref{eq:deltasin}) in terms of the parameters
$S$, $T$ and $U$. While the $U$-parameter vanishes in 
the approximation where only quartic terms or higher powers in $m_H$ 
are kept at the three-loop level, the $S$- and the $T$-parameters
give the following contributions
\begin{eqnarray}
&& S^{(3)} = \frac{1}{4\pi} {\left(\frac{g^2}{16\pi^2}\right)}^2
                   \frac{m_H^4}{M_W^4}
                   \left(\, 1.1105 \, \right) \,,\\
&& T^{(3)} =  \frac{1}{4\pi c_W^2} {\left(\frac{g^2}{16\pi^2}\right)}^2
                   \frac{m_H^4}{M_W^4}
                   \left(\, -1.7282 \, \right) \,.
\end{eqnarray}  
Using $\,g^2 \,=\, e^2/s_W^2 \,=\, 4\pi\alpha/s_W^2$ for the weak
coupling constant, with \mbox{$\alpha \,=\, 1/137$} and $s_W^2 \,=\, 0.23$,
the shifts to $\rho$, $\sintheff$ and $M_W$ are
\begin{eqnarray}
&& \Delta^{(3)}\rho =  -8.3\times 10^{-9}\times m_H^4/M_W^4\,,\\
&& \Delta^{(3)}\sintheff = 4.6\times 10^{-9}\times m_H^4/M_W^4 \,,\\
&& \Delta^{(3)} M_W  = -6.3\times 10^{-4}\mbox{MeV}\times m_H^4/M_W^4 .
\end{eqnarray}
\begin{figure}
\begin{center}
% \input{rhoplot_acta.tex}
%GNUPLOT: LaTeX picture with Postscript
\begin{picture}(0,0)%
\includegraphics{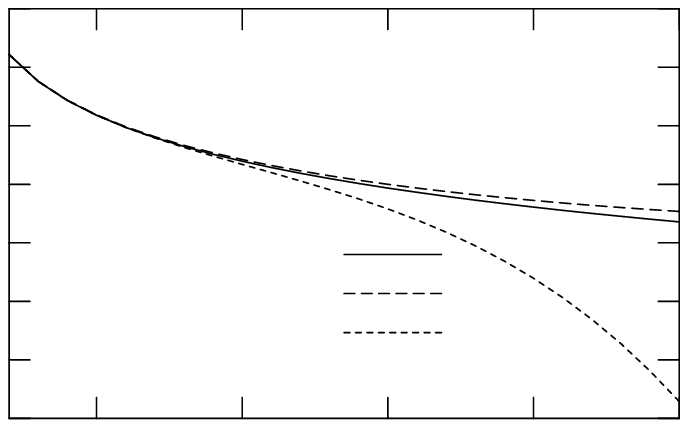}%
\end{picture}%
\setlength{\unitlength}{0.0200bp}%
\begin{picture}(13500,8100)(0,0)%
\put(2750,1650){\makebox(0,0)[r]{\strut{}-0.007}}%
\put(2750,2493){\makebox(0,0)[r]{\strut{}-0.006}}%
\put(2750,3336){\makebox(0,0)[r]{\strut{}-0.005}}%
\put(2750,4179){\makebox(0,0)[r]{\strut{}-0.004}}%
\put(2750,5021){\makebox(0,0)[r]{\strut{}-0.003}}%
\put(2750,5864){\makebox(0,0)[r]{\strut{}-0.002}}%
\put(2750,6707){\makebox(0,0)[r]{\strut{}-0.001}}%
\put(2750,7550){\makebox(0,0)[r]{\strut{}0.000}}%
\put(4284,1100){\makebox(0,0){\strut{} 5}}%
\put(6382,1100){\makebox(0,0){\strut{} 10}}%
\put(8479,1100){\makebox(0,0){\strut{} 15}}%
\put(10577,1100){\makebox(0,0){\strut{} 20}}%
\put(12675,1100){\makebox(0,0){\strut{} 25}}%
\put(550,4600){\rotatebox{90}{\makebox(0,0){\strut{}$\Delta \rho$}}}%
\put(7850,275){\makebox(0,0){\strut{}$m_H/M_W$}}%
\put(7575,4010){\makebox(0,0)[r]{\strut{}$\Delta^{(1)}$}}%
\put(7575,3448){\makebox(0,0)[r]{\strut{}$\Delta^{(1)}+\Delta^{(2)}$}}%
\put(7575,2886){\makebox(0,0)[r]{\strut{}$\Delta^{(1)}+\Delta^{(2)}+\Delta^{(3)}$}}%
\end{picture}%
% \endinput
\end{center}
\itcaption{
One-, two- and three-loop shifts in $\rho$ as a function of $m_H/M_W$.}
\label{fig:rho}
\end{figure}
\begin{figure}
\begin{center}
%\input{sinthshift_acta.tex}
%GNUPLOT: LaTeX picture with Postscript
\begin{picture}(0,0)%
\includegraphics{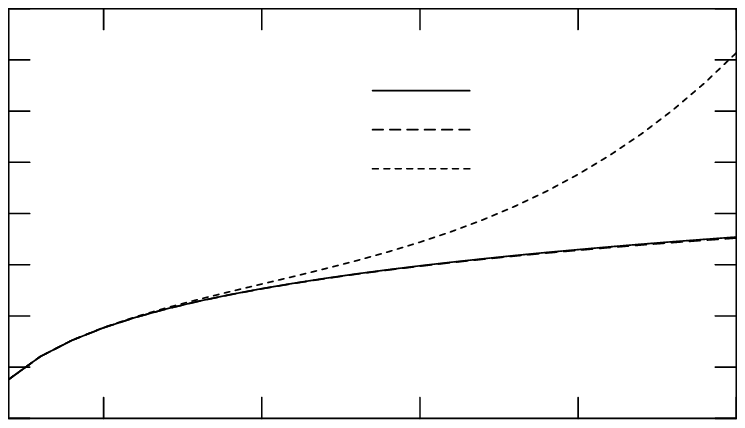}%
\end{picture}%
\setlength{\unitlength}{0.0200bp}%
\begin{picture}(13500,8100)(0,0)%
\put(1925,1650){\makebox(0,0)[r]{\strut{}0.0}}%
\put(1925,2388){\makebox(0,0)[r]{\strut{}0.5}}%
\put(1925,3125){\makebox(0,0)[r]{\strut{}1.0}}%
\put(1925,3863){\makebox(0,0)[r]{\strut{}1.5}}%
\put(1925,4600){\makebox(0,0)[r]{\strut{}2.0}}%
\put(1925,5338){\makebox(0,0)[r]{\strut{}2.5}}%
\put(1925,6075){\makebox(0,0)[r]{\strut{}3.0}}%
\put(1925,6813){\makebox(0,0)[r]{\strut{}3.5}}%
\put(1925,7550){\makebox(0,0)[r]{\strut{}4.0}}%
\put(3566,1100){\makebox(0,0){\strut{} 5}}%
\put(5843,1100){\makebox(0,0){\strut{} 10}}%
\put(8121,1100){\makebox(0,0){\strut{} 15}}%
\put(10398,1100){\makebox(0,0){\strut{} 20}}%
\put(12675,1100){\makebox(0,0){\strut{} 25}}%
\put(550,4600){\rotatebox{90}{\makebox(0,0){\strut{}$\Delta \sin^2\theta_{eff}\, (10^{-3})$}}}%
\put(7437,275){\makebox(0,0){\strut{}$m_H/M_W$}}%
\put(7163,6370){\makebox(0,0)[r]{\strut{}$\Delta^{(1)}$}}%
\put(7163,5808){\makebox(0,0)[r]{\strut{}$\Delta^{(1)}+\Delta^{(2)}$}}%
\put(7163,5246){\makebox(0,0)[r]{\strut{}$\Delta^{(1)}+\Delta^{(2)}+\Delta^{(3)}$}}%
\end{picture}%
% \endinput
\end{center}
\itcaption{
Shifts in $\sintheff$ as a function of $m_H/M_W$.}
\label{fig:sinth}
\end{figure}
The Higgs mass dependence of the $\rho$-parameter and
$\Delta\sintheff$ is shown in Figs.~\ref{fig:rho} and \ref{fig:sinth}.
It turns out that the sign of the leading three-loop corrections
to $\Delta\rho$, $\Delta\sintheff$ and $\Delta M_W$
is the same as the sign of the one-loop contributions.

The original question that motivated these calculations was, whether
inclusion of the three-loop corrections with strong interactions could
lead to an effect mimicking the one-loop effects of a light Higgs
boson. The result of the investigations shows that this is highly
unlikely. As the signs of the three-loop corrections are the same
as the ones of the one-loop corrections, with increasing Higgs mass,
the three-loop terms only make the effects grow faster, instead of
% the three-loop terms
partially cancelling the one-loop corrections. Therefore
the presence of a strongly interacting heavy Higgs-sector appears to
be extremely unlikely, and the electroweak precision data can indeed be
taken as a strong indication for a light Higgs boson sector.

\section*{Acknowledgement}
R.~B.~ would like to thank the organizers of the conference
for the pleasant atmosphere.
This work was supported by the Sofja Kovalevskaja Award
of the Alexander von Humboldt Foundation sponsored
by the German Federal Ministry of Education and Research and
the DFG within the project "(Nicht)-perturbative Quantenfeldtheorie".

\end{document}